\documentclass[aps,pra,twocolumn,groupedaddress,superscriptaddress]{revtex4}
\usepackage{amsmath}\usepackage{xcolor}\usepackage{epstopdf}
\usepackage{bbold}
\usepackage{graphicx}
\usepackage{color,ulem} 

\usepackage{natbib}
\usepackage{xcolor}

\newcommand{\Rfaqat}[1]{\textcolor{blue}{\fbox{RA} {\sl#1}}}

\newcommand{\be}{\begin{equation}}
\newcommand{\ee}{\end{equation}}
\newcommand{\beq}{\begin{eqnarray}}
\newcommand{\eeq}{\end{eqnarray}}
\newcommand{\bea}{\begin{array}}
\newcommand{\eea}{\end{array}}

\draft 
\begin{document}
\title{Lighting of a monochromatic scatterer with virtual gain  } 
\author{R. Ali}
\email[]{rali.physicist@gmail.com}

\affiliation{%
Applied Physics Department, Gleb Wataghin Physics Institute, University of Campinas, Campinas 13083-859, SP,
Brazil
}

\date{\today}

\begin{abstract}
In this work, we discuss the scattering features of a dipolar particle made of large refractive index material by employing the concept of virtual gain and virtual loss.  The virtual gain and loss  can be achieved in a lossless passive nanostructure by shaping the temporal waveform of incident signals in the complex frequency plane. We show that an appropriate tuning of excitation time of the impinging field allows to capture and release the electromagnetic energy on-demand for an arbitrary time scale in a lossless nanosphere. Thus, the nanosphere obliges to emit monochromatic magnetic light which can be tuned throughout the whole visible spectrum by varying the size of the nanosphere. This proposal may find fruitful applications  in lab-on-a-chip technologies and the realization of monochromatic sectoral multipole light source  with a large quality factor at nanoscale level. 
\end{abstract}

\maketitle 


\section{Introduction} 
Scattering of light by nanosphere is a core problem in electromagnetism which is thoroughly discussed in many applications such as sensing \cite{Kabashin2009,Ali2020},  
light manipulating devices  \cite{Kuznetsov2016,Almeida2004,Taghinejad2019}, optical manipulation  \cite{Ashkin1986,Ali2020c,Ali2021} and light emitting devices \cite{Kuznetsov2012}.
 Particularly, the nano-optical devices made of large refractive index nano-structures utilize the optical scattering for routing the optical fields at nanoscales level which can be efficiently controlled via electric and magnetic resonances \cite{Kuznetsov2012,Ali2020b}.
 
Remarkably, nanosphere made of semiconductors ubiquitous unique optical scattering properties due to strong magnetic response appears owing to sharp magnetic dipolar resonances with comparatively low losses in visible and near-infrared frequency-domain \cite{Evlyukhin2012,Dominguez2019}. The theoretical and experimental results predict that interaction of the impinging light with a nanosphere of large refractive index  $({\rm m_s} > 3${  e.g., of silicon, germanium, tin, gallium arsenide, molybdenum diselenide, lead selenide, mercury sulfide and so on)} rotates the displacement currents inside the sphere and yielding a magnetic moment oscillating through the center of the sphere  \cite{Person2013,fu2013,Etxarri2011,geffrin2012,alaee2015,Medina2011}. Thus,  a  strong magnetic response has appeared in a nonmagnetic sphere in the visible spectral range \cite{Evlyukhin2012}. 
 Therefore, they are considered as basic elements for the designing of complex metamaterials with desired optical properties such as optical clocking \cite{Farhat2012}, single particles Mie nanolaser \cite{Mylnikov2020}, and  Huygen source \cite{Kuznetsov2012,Ali2020b}.
  
One of the important applications of such resonators is an all-dielectric Mie nano-laser which utilizes a gain functionalized  dielectric nanostructure with a single lasing mode at room temperature \cite{Mylnikov2020,Tiguntseva2020,Hoang2020}.   A typical Mie-nanolaser starts lasing  a coherent monochromatic light at the certain threshold gain at which the enhancement of stimulated emission is achieved.
 It is well established that magnetic resonance supported lasing mode can be accomplished at a lower lasing threshold and large quality Q factor \cite{Kuznetsov2012,Zhang2018}. Therefore, a small optical gain is enough to overcome the radiative losses and increase its Q factor several times as compared to a regular Mie mode.
 Since the lasing modes are associated with Mie resonances of the resonator, therefore, they can be tuned by changing the dimensions of the particle,  structural properties  and amount of optical gain.

In this article, we provide an alternative method to design an all-passive and all-dielectric monochromatic lighting source without employing optical gain. In this context, we consider Silicon nano-sphere illuminated by non-monochromatic incident signal oscillating in complex frequency plane. 
{  Under the time-harmonic convention $e^{-i\omega t},$ the scattering zeros and scattering poles are located in the upper and lower frequency half plane, respectively. For the case of monochromatic incidence, a lossless cavity cannot admit the scattering zeros and scattering poles but  they can be reached by adding appropriate material losses and gain,  respectively, to the cavity. On the contrary,  for non-monochromatic incident signal oscillating in the complex plane, the cavity admits the scattering zeros and poles  which can be pushed  towards the real axis by tailoring the temporal profile of the incident field.  This temporal modulation of incident signal allows matching the exponentially diverging perfectly absorbing mode corresponding to the scattering  zero  over   a  finite  time interval. To this end, the scattering signal has to be zero. Since the cavity is lossless  (Ohmic losses are  zero), thus, the energy associated with the incident signal will be virtually absorbed or stored in the cavity.
  }

 {  On the other hand,  this stored energy can be released on demand by  cutting  the exponential growth of the impinging signal. Thus,  the cavity is obliged to give a strong scattering signal owing to release of the stored energy in the absence of gain. In a nutshell,} { the transient behavior of the incident signal dictates the lossless cavity to captures the incident signals for a finite time scale and release them on demand.} This concept of the capturing and releasing of the incident signal are known as virtual absorption and virtual gain, respectively, which has been recently proposed and experimentally demonstrated by Andrea  Al\' u$'$s  group \cite{Baranov2017a,Lepeshov2020,Radi2020,Li2020,Baranov2017}. 
 
 { Virtual loss and virtual gain in a lossless cavity can be achieved in the laboratory  by exciting the cavity with multiple incident signals with controllable relative intensities  and relative phases in real time. The spatiotemporal modulation of the input  waveforms  is then used to store and release the electromagnetic energy at will \cite{Baranov2017}.  In this way, the cavity } can engage the scattering zeros  in the upper complex frequency half-plane when the incident signal grows faster than decay. On the contrary, the virtual gain can be achieved by engaging the scattering poles positioned in the lower complex frequency half plane, where the incident signal grows weaker than its decay rate. Thus, the cavity is obliged to store energy in the upper and release energy in the lower frequency plane.  {  This counterintuitive phenomenon plays crucial role in many applications, i. e. coherent perfect virtual  absorption \cite{Baranov2017,Longhi2018}, critical coupling of the signals to a high-Q lossless resonators \cite{Radi2020} and tuning the direction of radiation pressure  \cite{Lepeshov2020}}.
 
 In this work, we use the {   well established concepts and findings displayed in the Refs. \cite{Kuznetsov2012,Person2013,fu2013,Etxarri2011,geffrin2012,alaee2015,Medina2011,Baranov2017a,Lepeshov2020,Radi2020,Li2020,Baranov2017} to achieve, for the first time,} all passive monochromatic nano-emitter by pairing the Mie resonances with the scattering poles in a balanced way,  {which allows to emit a three orders magnitudes large  monochromatic signal at a nanoscale level.} 

{ The rest of the  paper is organized as follows. Section \ref{method} is devoted to the  brief description of  methodology, where the scattering cross sections are  discussed using the  Mie scattering theory. In Sec. \ref{results}, we present numerical analysis  and results are discussed. Finally, we
summarize  our results and conclude the findings  in Sec. \ref{summary}.}

\section{Methodology} \label{method}
 We start  by considering a  plane wave scattering by a nanoparticle of radius $R$,  refractive index ${\rm m_s}$ suspended in a nonabsorbing medium with refractive index ${\rm n_m}$. The incident time-harmonic optical field of angular frequency $\omega$ is propagating along the z-axis,  polarization along the x-axis and impinging on the nanosphere placed at the origin.
   The electromagnetic self-oscillation of the nanoparticle can be determined by  following the standard approach presented in Ref. \cite{Bohren}.
{ Here, the incident electromagnetic fields are expressed in spherical coordinates in terms of  vector spherical harmonics  and the corresponding incident field  ${\bf E_{in}}$, scattered  field ${\bf E}_{s}$  and field inside ${\bf E}_{int}$ the sphere are  given  as (see Ch. 4 of the Ref. \cite{Bohren}) 
\begin{equation}
{\bf E}_{in} = E_0 \sum^{\infty}_{\ell =1} i^\ell \frac{2\ell+1}{\ell(\ell+1)}(M^{(1)}_{o1\ell} - i N^{(1)}_{e1\ell}),
\end{equation}
\begin{equation}
{\bf E}_{s} = E_0 \sum^{\infty}_{\ell =1} i^\ell \frac{2\ell+1}{\ell(\ell+1)}( i a_\ell N^{(3)}_{e1\ell} - b_\ell M^{(3)}_{o1\ell} ),
\end{equation}
\begin{equation}
{\bf E}_{int} = E_0 \sum^{\infty}_{\ell =1} i^\ell \frac{2\ell+1}{\ell(\ell+1)}(  c_\ell M^{(1)}_{o1\ell} - i d_\ell M^{(1)}_{e1\ell}),
\end{equation}}
{ where, ${\bf M}_{e1\ell}$,\,  ${\bf M}_{o1\ell}$, \,  ${\bf N }_{e1\ell}$ and  ${\bf N }_{o1\ell}$ are the  vector
spherical harmonics, and  the explicit expression of these vector functions are present in Ref. \cite{Bohren}. In addition, $a_\ell$, \, $b_\ell$\, and  $c_\ell$,\, $d_\ell$ are the sets of the Mie coefficients representing the scattered electromagnetic field amplitudes inside the surrounding medium  and field amplitudes  inside the spherical particle, respectively. These Mie coefficients can be determined by applying the subsidiary boundary conditions at the surface of the sphere $(r=R)$.  {For instance, the  electric  and magnetic fields must obey the following boundary conditions at the boundary between the sphere and the  surrounding medium  } \cite{Bohren} }

{
\begin{equation}
({\bf E}_{in} + {\bf E}_{s} - {\bf E}_{int}) \times \hat{n} =0, \label{BC1}
\end{equation}
 \begin{equation}
({\bf H}_{in} + {\bf H}_{s} - {\bf H}_{int}) \times \hat{n} =0. \label{BC2}
\end{equation}}

{  Finally, the quantification of  energy received and scattered  by the sphere is defined in terms of dimensionless quantities} and  can be expressed  as extinction efficiency $Q_{ext}$, scattering efficiency  $Q_s$,  and absorption  efficiency $Q_{abs}$  \cite{Ali2020c,Bohren,Ali2018} as

\begin{figure}
\centering
\includegraphics[width =3.4 in]{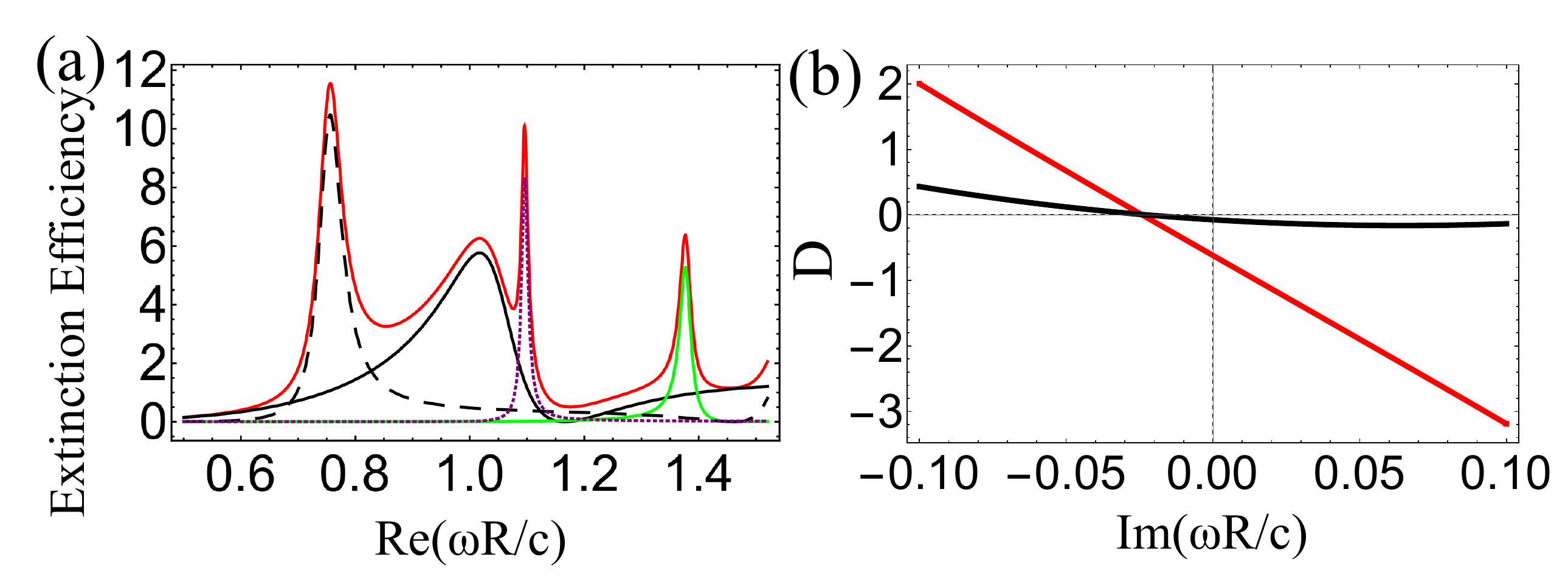}
\caption{ (a)  Normalized extinction  efficiency (red) by a sphere of radius ${\rm R}$  as a function of normalized real frequency. The scattering contribution of a magnetic dipole  (dashed), electric dipole  (black),  magnetic quadruple (dotted), and electric quadruple (green).
(b) Real (red) and imaginary (black) part of denominator ${\rm D}$ of magnetic dipolar Mie-coefficient  as a function of normalized imaginary frequency for fixed real frequency at 0.75.} 
 \label{F1}
\end{figure}

\begin{equation}
Q_{ext} = \frac{2}{x^2} \sum^\infty_{\ell=1} (2\ell+1)Re(a_\ell+ b_\ell), \label{qext}
\end{equation} 

\begin{equation}
Q_{s} = \frac{2}{x^2} \sum^\infty_{\ell=1} (2\ell+1)(|a_\ell|^2+|b_\ell|^2), \label{qsca}
\end{equation} 

\begin{equation}
Q_{abs} = Q_{ext} -Q_{s}  \label{qabs}.
\end{equation}

where, $ x = ka$ ($k=2\pi/\lambda$) is the  size parameter, and $\lambda$ is the incident wavelength.
{  The scattering  Mie coefficients $a_\ell$ and $b_\ell$ are  representing  the scattering amplitudes of the electric and  magnetic field, respectively,  which can easily  be  determined by simplification of the  Eqs. \ref{BC1} \&  \ref{BC2} and one  can read as \cite{Bohren,Ali2018}

\begin{eqnarray}
a_{\ell}= \frac{m\psi_{\ell}(mx) \psi'_{\ell}(x)-\mu\psi_{\ell}(x) \psi'_{\ell}(mx)}{m\psi_{\ell}(mx) \xi'_\ell(x)-\mu\xi_\ell (x) \psi'_{\ell}(mx)},  \label{Mie_Ca}
 \end{eqnarray}
 \begin{eqnarray}
  b_{\ell}=  \frac{\mu \psi_{\ell}(mx) \psi'_{\ell}(x)-m  \psi_{\ell}(x) \psi'_{\ell}(mx) }{\mu \psi_{\ell}(mx) \xi'_\ell(x)-m\xi_\ell(x)\psi'_{\ell}(mx) }, \label{Mie_Cb} 
 \end{eqnarray}
 
where, $\psi_\ell$, $ \xi_\ell $ are Riccati-Bessel functions \cite{DLMF25.12}, ${\rm m = m_s /n_m}$ and ${\rm \mu = \mu_s /\mu_m}$ are the relative refractive index and relative permeability, respectively.
The index $\ell$ denotes the $\ell^{th}$ order of spherical harmonics. }
It is worthy to emphasize  that for monochromatic incident light, a losses cavity presents $Q_{abs}= 0$ and all energy received by the sphere is being scattered and allowing $Q_{ext} =Q_{s}$. However, for the case of non-monochromatic incidence, the situation dramatically changes and a passive cavity presents nonzero $Q_{abs}$ without heat losses. 

\begin{figure}
\centering
\includegraphics[width =3.4 in]{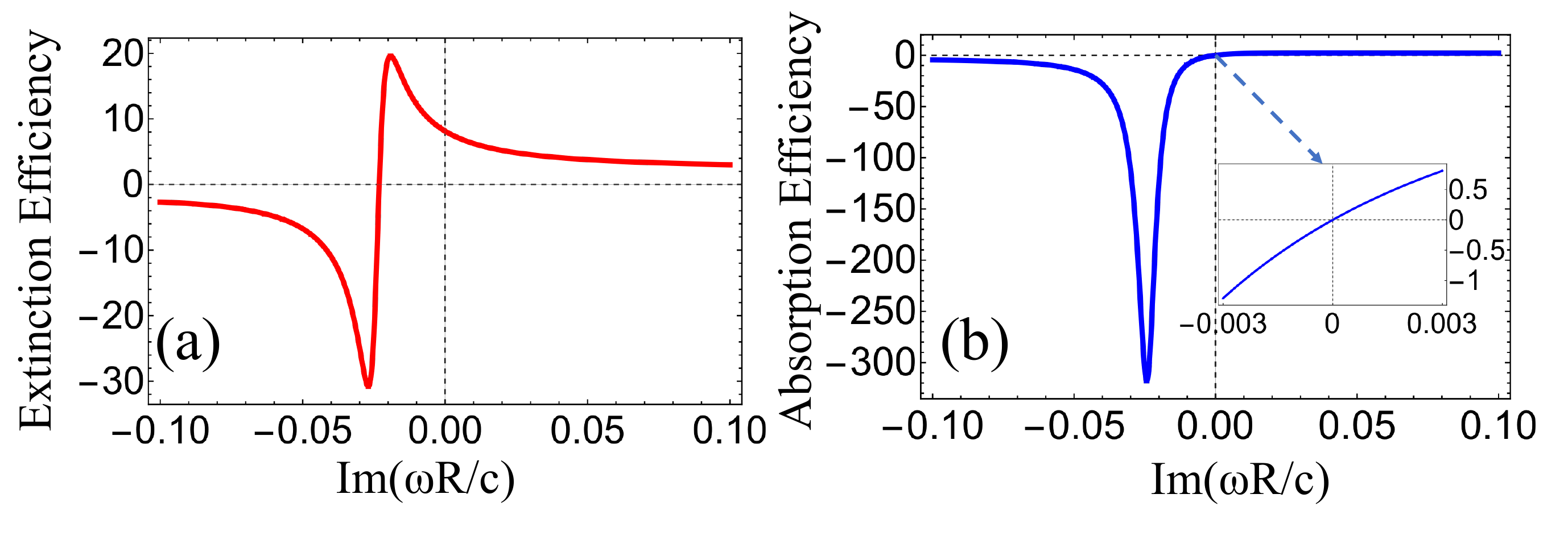}
\caption{ Normalized (a) extinction efficiency and (b) absorption efficiency by a sphere of radius ${\rm R}$  illuminated by non-monochromatic incident field as a function of normalized imaginary frequency for fixed real frequency at at fixed ${\rm Re(\omega R/c = 0.75 )}$} 
 \label{F2}
\end{figure}

\section{Results and discussion} \label{results}

In the following discussion, we consider  two-port non-monochromatic light source $E^{in}(t) \propto E_0 e^{-i\omega t}$ incident upon a nanoparticle  of radius  $R$ and refractive index ${\rm m_s= 4}$, where $\omega= \omega'+i \omega''$,  $\omega'$  and  $\omega'' $  are the real and imaginary frequencies, respectively.  Under this time convention, a passive sphere can engage the scattering zero for $\omega'' >0$  and scattering poles for  $\omega'' < 0$ in the complex frequency plane.  As the incident signal is oscillating in the complex frequency plane,  the position of the scattering zero and scattering poles can be conveniently tailored by tuning the time evolution of complex excitation of the incident field, instead of using optical losses or optical gain.
  Thus, tailoring of the temporal wavefront of incident signal one can push the scattering zeros and scattering poles close to the real frequency axis over a finite time interval.
For instance, in the upper complex frequency plane, the  incident signal is growing exponentially  $E^{in}(t)\propto E_0 e^{|\omega'' |t}$, as the complex frequency increases the lossless cavity is obliged to exhibit  $Q_{ext} > Q_s $ and at some points the scattering output approaches to zero. Hence, the scattering zero has been reached and all incident energy is being stored in the lossless cavity in terms of so-called virtual loss (without Ohmic losses)  in the resonator. On the contrary, in the lower complex frequency plane  incident signal $E^{in}(t)\propto E_0 e^{-|\omega'' |t}$ exponentially attenuates, where the signal decays faster than its growth rate, and hence the particle scatters more energy than it receives $Q_{ext} <Q_s $. Since the particle has no gain therefore  the additional scattered energy is coming from the stored energy  \cite{Trainiti2019}.
{ In this context, the  scattering signal will be much stronger, than the ordinary Mie scattering mode calculated at the real frequency axis,  and monochromatic close to the complex poles due to virtual gain, which is the main goal of this work.}

In order to make discussion independent of the incident frequency range,  we calculate the dimensionless scattering cross sections as a function of dimensionless frequency ${\rm \omega R/c}$, where c is the speed of light.  Figure \ref{F1}(a) demonstrates the normalized scattering efficiency as a function of normalized monochromatic frequency (at the real frequency axis).  It is worthwhile  to note that the sphere is nonmagnetic and having large refractive index, the incident field rotates the displacement currents and providing a strong ring-shaped electric field inside the sphere. Consequently, a strong transversely oscillating magnetic field in the center of the ring appears at $\lambda_0/m_s \approx 2R$ and providing an oscillating magnetic dipole (MD) at ${\rm Re( \omega R/c)=0.75}$ \cite{Kuznetsov2016}. { It is important to emphasize  that the sizes of the nanosphere ($R \approx  \lambda_0/2m_s $) required to excite the dipolar resonance are sufficiently large and hence the nonlocal and quantum effects can be ignored \cite{Ruppin1973,Gubbin2020}.  } As  frequency grows other  resonances corresponding to electric dipoles (ED), magnetic quadruple (MQ) and electric quadrupole (EQ  occur at  ${\rm Re( \omega R/c) = 1}$, ${\rm Re( \omega R/c)=1.1}$,  and  ${\rm Re(\omega R/c) = 1.4}$, respectively, as expected \cite{fu2013,Etxarri2011,geffrin2012,Medina2011}. 
It is clearly shown that in the long-wavelength regime the fundamental resonance MD has larger Q factor than  ED. Thus,  MD should be a good candidate for the scattering of magnetic light for longer time scale.
 It is worthy to stress that sphere is lossless and hence $Q_{s} = Q_{ext}$

\begin{figure}
\centering
\includegraphics[width =3.4 in]{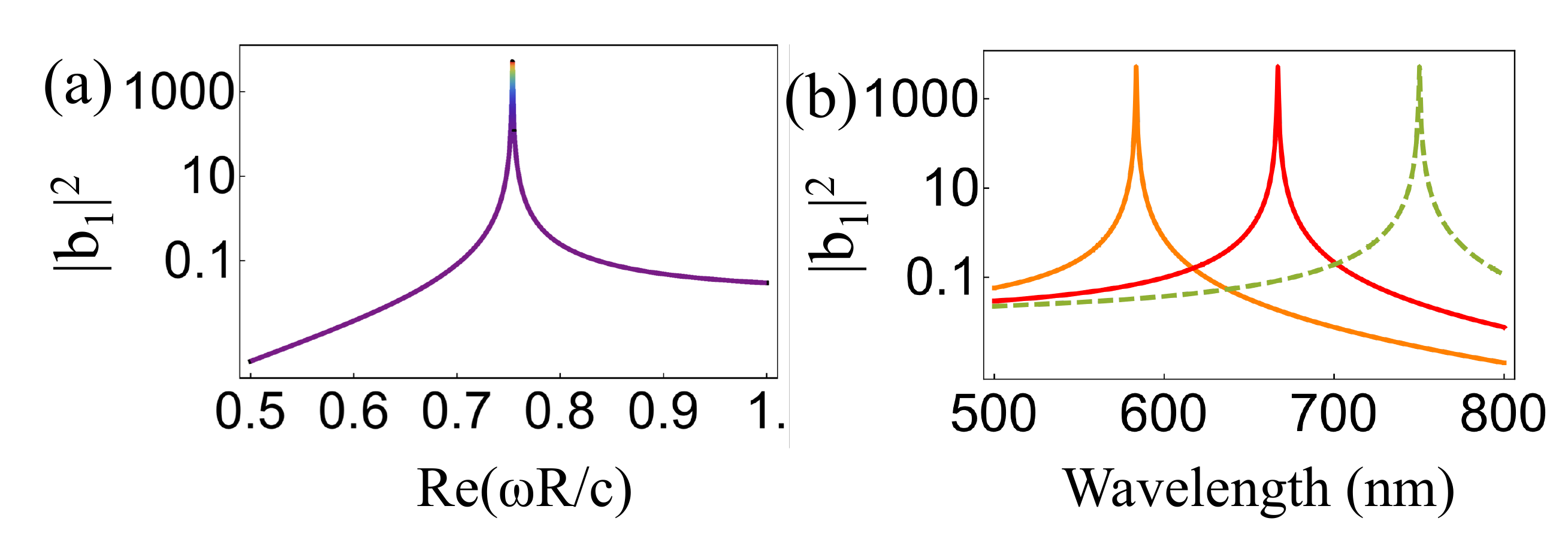}
\caption{ (a)  The  squared absolute value of the dipolar magnetic Mie coefficient $|b_1|^2$  as a function of normalized  real frequency.  (b)  The scattered frequency shift in the visible region, where the scattering response of  $|b_1|^2$ is calculated  by considering a sphere of radii ${\rm 70 \, nm}$ (orange), ${\rm 80 \, nm}$ (red) and ${\rm 90 \, nm}$ (dashed) as a function of wavelength. The normalized  imaginary frequency is fixed for all calculations at ${\rm Im(\omega R/c) = -0.02}$.} 
 \label{F3}
\end{figure}

 On the other hand,  if we shin a non-monochromatic field the situation drastically changes and the lossless cavity leaves nonzero absorption footprint in the complex frequency plane.
 The theoretical study predicts that the real frequencies corresponding to the resonances should engage scattering poles in the lower frequency plane. The position of such poles can be analytically calculated by solving the denominator of corresponding scattering amplitudes. As we are interested in a magnetic  monochromatic dipolar  light source, therefore, we numerically solve denominator of magnetic dipolar ($\ell=1$) Mie coefficient $D={\mu \psi_{\ell}(mx) \xi'_\ell(x)-m\xi_\ell(x)\psi'_{\ell}(mx) }$ along the complex frequency axis for ${\rm Re(\omega R/c) = 0.75}$ and shown in Fig. \ref{F1}(b).  It is straightforward to notice that   ${ \rm Re(D)}$ (red) and ${\rm Im(D)}$ (black) are vanishing at  ${\rm Im( \omega R/c)=-0.02}$, and it is a clear indication that a pole of $b_1$ in complex frequency plane has been reached. In order to discuss the scattering behavior  in complex frequency plane close to the pole, we calculate the extinction  efficiency Fig. \ref{F2}(a) and abortion efficiency Fig. \ref{F2}(b) as a function of imaginary frequency for   ${\rm Re( \omega R/c) = 0.75}$.
 
It is important to stress that a  resonator engages corresponding complex zero in the upper complex frequency plane and captures energy (virtual loss) during the transient excitation. This stored energy can be paired with poles and released on will as depicted in Fig. \ref{F2}, where the particle has to decay incident signal faster than it grows and hence scattering efficiency should  be different than the extinction efficiency (for $\omega'' \neq 0$).  It is clearly shown in Fig. \ref{F2}(a) that when magnetic dipolar resonance  is  aligned with scattering pole  positioned  at ${\rm Im(\omega R/c) = - 0.02}$  
and the  resonator scatters more energy than it receives, thus negative extinction efficiency is appearing. Hence it is immediate to tell that the rest part of the scattering energy is coming from the virtual gain which  is  the consequence of the negative absorption for $\omega'' < 0$ as shown in Fig. \ref{F2}(b). This figure shows that the  absorption is zero at the real frequency axis while $Q_{abs} > 0$ and  $Q_{abs} < 0$ in the upper and lower frequency plane, respectively, as one can refer to the inset of  Fig. \ref{F2}(b). It is worth mentioning  that the phenomenon of perfect energy capture and release at will has been demonstrated experimentally \cite{Baranov2017,Trainiti2019}. 

\begin{figure}
\centering
\includegraphics[width =3.4 in]{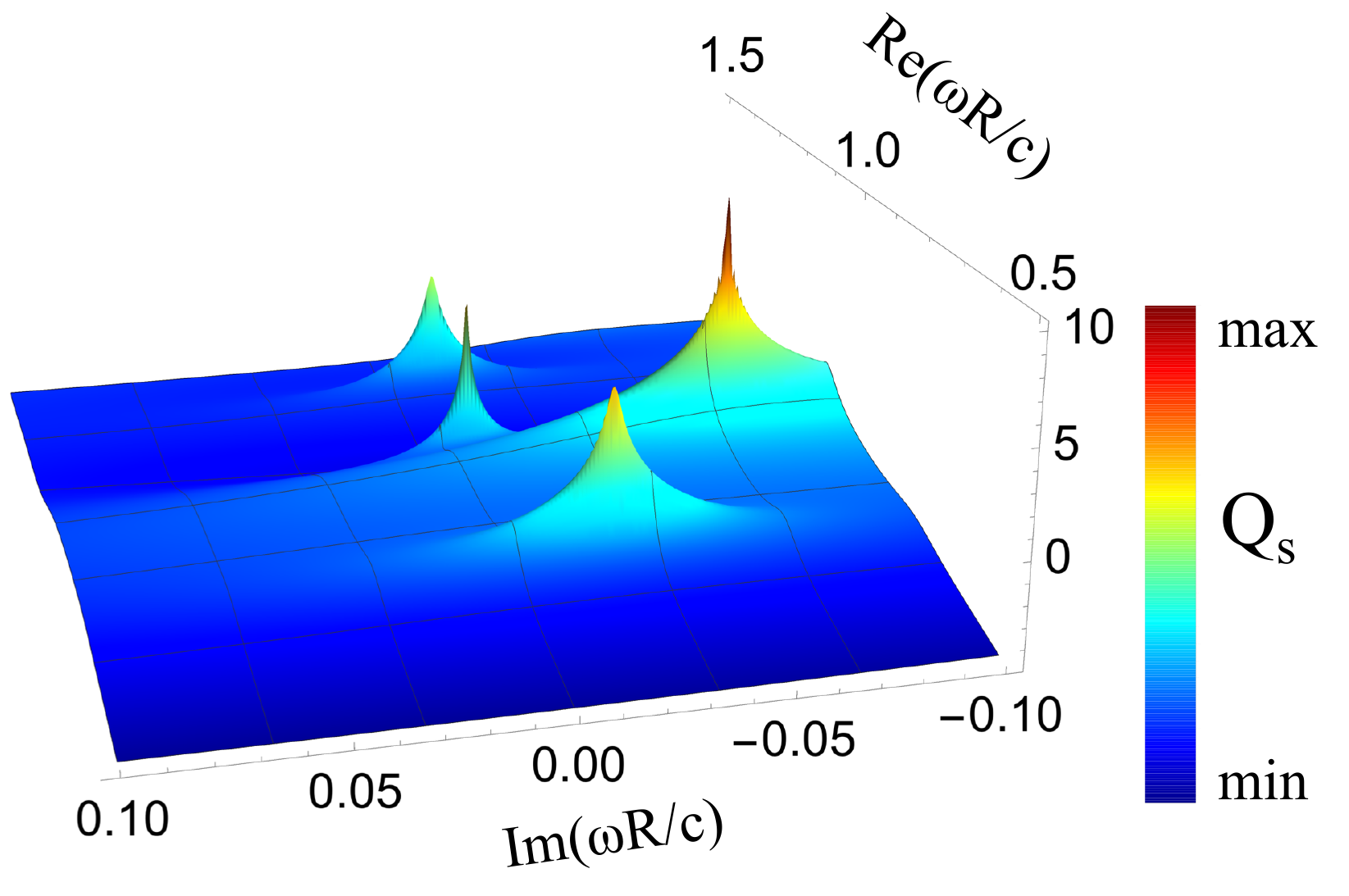}
\caption{Density plot showing the scattering pattern of light by a sphere of radius R illuminated by a non-monochromatic incident signal as a function of real and imaginary frequency.   } 
 \label{F4}
\end{figure}

Figure \ref{F3}(a) demonstrates the squared absolute value of the magnetic dipole  as a function of  ${\rm Re(\omega R/c) }$ for  ${\rm Im = -0.02}$.  It is shown  that the  magnetic dipolar scatterer behaves like a monochromatic magnetic light source of large output signal. The scattering signal is about three order magnitude stronger than the value achieved for  monochromatic incidence as shown in Fig. \ref{F1}(a).  Such type of scattering phenomenon can also be achieved by employing an optical gain medium \cite{Mylnikov2020}. 
The scattered magnetic light by the nanoparticle is so strong and it can be easily seen under a dark-field optical microscope.  It is important  to emphasize that the wavelength of the monochromatic scattered magnetic light signal can be varied throughout the whole visible spectral range from violet to red by just changing by changing the cavity size and refractive index.  For concreteness, we validate our results by considering the realistic  material and physical  parameters, and the corresponding scattering signal is 
 demonstrated in Fig. \ref{F3}(b). Here we present squared absolute value of the magnetic dipole $|b_1|^2$ of a Si nanosphere as a function of wavelength for different  sizes: ${ \rm R= 70\, nm}$ (orange), ${ \rm R= 80\, nm}$ (red) and ${ \rm R= 90\, nm}$ (dashed).
 It is clearly shown that for a small increment  in the size of the nanoparticle the resonance peak moves from
 ${\rm 480 \, nm}$ 
  to ${\rm 700\, nm}$.   
Finally, we expand this concept to parametric space to elucidate the role of other resonances by plotting  the scattering amplitudes as a function of normalized real and imaginary frequency in Fig. \ref{F4}. It is clearly shown that there are four scattering peaks corresponding to MD, ED,  EQ, and  MQ. Since we are interested in a smallest scatterer, therefore, we are considering the MD dipolar resonance for dipolar magnetic light. On the other hand, the ED signal appears with a low Q factor and the MD and ED would not be important  in the long wavelength or dipolar limit. 

\section{Summary} \label{summary}

In this paper, we have revisited the scattering  response  of  non-monochromatic incident  light, oscillating with the complex frequency, by a large refractive index nanosphere  using the  Mie scattering theory.
Our numerical analysis shows that one can achieve  monochromatic magnetic light with a large quality factor than an ordinary Mie mode without employing optical gain and population inversion at room temperature.  {  To this end, the monochromatic scattering signal  is achieved by spatiotemporal modulation of the   exponentially growing incident signal}. 
 Thus,  fine tailoring of the excitation in the complex frequency plane provides an optimal  control to tune the location of the scattering poles and move them close to the real frequency axis, {  instead  of using optical gain which was considered to be an essential requirement.  Our findings have shown that the efficient tuning of the incident signal's  excitation in complex frequency plane allows to pair the scattering poles and scattering zero in a balanced way. Thus, } providing  an opportunity to realize monochromatic  dipolar electric, magnetic light  \cite{Trigo2019}. This smallest Mie-resonant is able to emit single-valued wavelength light which can be tuned throughout the visible region by varying the size of the nanosphere. Finally, we confident that this novel scattering paradigm can pave the way to design the multipolar light source, multifunctional photonic designs for routing light at the nanoscale level.

\subsection*{Acknowledgment} We thank S. Iqbal, G. Wiederhecker, F. A. Pinheiro, F. S. S. Rosa and P. A. M. Neto for inspiring discussions. This work is partially supported by 
 Fundac\~ao de Amparo a   Pesquisa do Estado de S\~ao Paulo  (FAPESP) (2020/03131-2). 
The author declares no conflicts of interest.


\end{document}